\documentclass[runningheads,envcountsame,a4paper]{llncs}

\usepackage{amsmath} 
\usepackage{amssymb} 
\usepackage{booktabs} 
\usepackage[caption=false]{subfig} 
\usepackage[hyphens]{url} 
\usepackage{listings} 
\usepackage{microtype} 
\usepackage{multirow} 
\usepackage{pgfplots} 
\pgfplotsset{width=7cm,compat=1.14}
\usepackage{threeparttable} 
\usepackage{tikz} 
\usetikzlibrary{patterns,shapes}

\newcommand{\highlight}[1]{\textbf{#1}}
\newcommand{\appar}[1]{\emph{#1\hspace{0.25ex plus 0.25ex}}}
\newcommand{\itemLabel}[1]{\item[\emph{#1}]}
\newcommand{\dslKey}[1]{\underline{\emph{#1}}}

\newcommand{\vect}[1]{\mathbf{#1}}
\newcommand{\y}{\vect{y}}
\newcommand{\Y}{\vect{Y}}
\newcommand{\f}{\vect{f}}
\newcommand{\F}{\vect{F}}

\newcommand{\yk}{\y_\kappa}
\newcommand{\tk}{t_\kappa}
\newcommand{\hk}{h_\kappa}
\newcommand{\R}{\mathbb{R}}

\newcommand{\offline}{off\-{\hspace{0pt}}line}
\newcommand{\Offline}{Off\-{\hspace{0pt}}line}

\newenvironment{customlegend}[1][]{%
  \begingroup
  \csname pgfplots@init@cleared@structures\endcsname
    \pgfplotsset{#1}%
  }{%
    \csname pgfplots@createlegend\endcsname
    \endgroup
  }%
  \def\addlegendimage{\csname pgfplots@addlegendimage\endcsname}

\begin{document}


\title{Offsite Autotuning Approach}
\subtitle{Performance Model Driven Autotuning Applied to Parallel Explicit ODE 
Methods}
\titlerunning{Offsite -- A Performance Model Driven Autotuning Approach}



\author{Johannes Seiferth, Matthias Korch, Thomas Rauber}

\institute{
  Department of Computer Science, University of Bayreuth, Bayreuth, Germany
  \email{$\{$johannes.seiferth, korch, rauber$\}$@uni-bayreuth.de}
}


\maketitle


\begin{abstract}
Autotuning (AT) is a promising concept to minimize the often tedious manual 
effort of optimizing scientific application for a specific target platform. 
Ideally, an AT approach can reliably identify the most efficient implementation 
variant(s) for a new platform or new characteristics of the input by applying 
suitable program transformations and analytic models.
In this work, we introduce Offsite, an \offline~AT approach which automates this
selection process at installation time by rating implementation variants based 
on an analytic performance model without requiring time-consuming runtime tests.
From abstract multilevel YAML description languages, Offsite automatically 
derives optimized, platform-specific and problem-specific code of possible 
variants and applies the performance model to these variants.

We apply Offsite to parallel numerical methods for ordinary differential 
equations (ODEs). In particular, we investigate tuning a specific class of 
explicit ODE solvers (PIRK methods) for four different initial value problems 
(IVPs) on three different shared-memory systems. Our experiments demonstrate 
that Offsite can reliably identify the set of most efficient implementation 
variants for different given test configurations (ODE solver, IVP, platform) and
effectively handle important AT scenarios.

\end{abstract}


\section{Introduction}
\label{sec:introduction}


The performance of scientific applications strongly depends on the 
characteristics of the targeted computing platform, such as, e.g., the processor
design, the core topology, the cache architectures, the memory latency or the 
memory bandwidth. Facing the growing diversity and complexity of today's 
computing landscape, the task of writing and maintaining highly efficient 
application code is getting more and more cumbersome for software developers. A 
highly optimized implementation variant on one target platform, might, however, 
perform poorly on another platform. That particular poorly performant 
implementation variant, though, could again potentially outperform all other 
variants on the next platform. Hence, in order to achieve a high efficiency and 
obtain optimal performance when migrating an existing scientific application, 
developers need to tune and adapt the application code for each specific 
platform anew.


\subsection{State of the Art}
\label{sec:related_work}


A promising concept to avoid this time-consuming, manual effort is 
\highlight{autotuning} (AT), and many different approaches have been proposed to
automatically tune software \cite{Balaprakash2018}. AT is based on two core 
concepts: \emph{(i)} the generation of optimized implementation variants based 
on program transformation and optimization techniques such as, e.g., loop 
unrolling or loop tiling, and \emph{(ii)} the selection of the most efficient 
variant(s) on the target platform from the set of generated variants.
In general, there are \emph{(i)} \highlight{\offline} and \emph{(ii)} 
\highlight{online} AT techniques. \Offline~AT tries to select the supposedly 
most efficient variant at compile or installation time without actual knowledge 
of the input data. These approaches are applicable for use-cases, whose 
execution behavior does no depend on the input data. This is the case, e.g., for
dense linear algebra problems, which can, i.a., be tuned \offline~with 
\emph{ATLAS} \cite{Whaley2001} and \emph{PhiPAC} \cite{Bilmes1997}. In other 
fields, such as sparse linear algebra or particle codes, characteristics of the 
input data heavily influence the execution behavior. By choosing the best 
variant at runtime---when all input is known---, online AT approaches such as 
\emph{Active Harmony} \cite{Tiwari2011} and \emph{ATF} \cite{Rasch2019} 
incorporate these influences.

Selecting a suitable implementation variant from a potentially large set of 
available variants in a time-efficient manner is a big challenge in AT. Various 
techniques and search strategies have been proposed in previous works to meet 
this challenge \cite{Balaprakash2018}. A straightforward approach is the 
time-consuming comparison of variants by runtime tests, possibly steered by a 
single search strategy, such as an exhaustive search or more sophisticated 
mathematical optimization methods like \emph{differential evolution} 
\cite{Das2016}, or a combination of multiple search strategies \cite{Ansel2014}.
\cite{Pfaffe2019} proposes a hierarchical approach that allows the use of 
individual search algorithms for dependent subspaces of the search space. As an 
alternative to runtime tests, analytic performance models can be applied to 
either select the most efficient variant or to reduce the number of tests 
required by filtering out inefficient variants beforehand.
In general, two categories of performance models are distinguished: 
(i)~\highlight{black box models} applying statistical methods and machine 
learning techniques to observed performance data like hardware metrics or 
measured runtimes in order to learn to predict performance behavior 
\cite{Shudler2019,Tallent2009}, and (ii)~\highlight{white box models} such as 
the \emph{Roofline model} \cite{Williams2009} or the \emph{ECM performance 
model} \cite{Hofmann2019,Stengel2015} that describe the interaction of hardware 
and code using simplified machine models. For loop kernels, the \emph{Roofline} 
and the \emph{ECM model} can be automatically constructed with the 
\emph{Kerncraft} tool \cite{Hammer2016}.











\subsection{Main Contributions}
\label{sec:contributions}


In this work, we propose \highlight{Offsite}, an \offline~AT approach that 
automatically identifies the most efficient implementation variant(s) during 
installation time based on performance predictions. These predictions stem from 
an analytic performance prediction methodology for explicit ODE methods proposed
by \cite{Seiferth2018} that uses a combined white and black box model approach 
based on the ECM model. The main contributions of this paper are:

\appar{(i)}
We develop a novel \offline~AT approach for shared-memory systems based on performance modelling. This approach automates the task of generating the pool of possible implementation variants using abstract description languages. For all these variants, our approach can automatically predict their performance and identify the best variant(s). Further, we integrated a database interface for collected performance data which enables the reusability of data and which allows to include feedback from possible online AT or actual program runs.

\appar{(ii)}
We show how to apply Offsite to an algorithm from numerical analysis with complex runtime behavior: the parallel solution of IVPs of ODEs.

\appar{(iii)}
We validate the accuracy and efficiency of Offsite for different test configurations and discuss its applicability to four different AT scenarios.


\subsection{Outline}
\label{sec:outline}


Section \ref{sec:testbed} details the selected example use-case (PIRK methods) 
and the corresponding testbed. Based on this use-case, Offsite is described in 
Section \ref{sec:offsite}. In Section \ref{sec:experiments}, we experimentally 
evaluate Offsite in four different AT scenarios and on three different target 
platforms. Section \ref{sec:conclusion} concludes the paper.


\section{Use-Case and Experimental Test Bed}
\label{sec:testbed}


\subsection*{Use-Case: PIRK Methods}
\label{sec:testbed:pirk}


As example use-case, we investigate \highlight{parallel iterated Runge-Kutta} 
(PIRK). PIRK methods are part of the general class of explicit ODE methods 
\cite{VdHouwen1990} and solve an ODE System
\begin{equation}
  \y'(t) = \f(t, \y(t)) \enspace , \enspace \y(t_0) = \y_0 \enspace, \enspace \y \in \R^{n}
  \label{eq:ode}
\end{equation}
by performing a series of timesteps until the end of the integration interval 
is reached. In each timestep, a new numerical approximation $\y_{\kappa + 1}$ 
for the unknown solution $\y$ is determined by an explicit predictor--corrector 
process in a fixed number of substeps.


\begin{figure}[tbp]
  \centering
  \input{fig/listings/code_pirk.tex}
\end{figure}


PIRK methods are an excellent candidate class for AT. Their complex 
four-dimensional loop structure (Lst. \ref{lst:pirk}) can be modified by loop 
transformations resulting in a large pool of possible implementation variants 
whose performance behavior potentially varies highly depending on:
\emph{(i)}
the composition of computations and memory accesses,
\emph{(ii)}
the number of stages of the base ODE method,
\emph{(iii)}
the characteristics of the ODE system solved,
\emph{(iv)}
the target hardware,
\emph{(v)}
the compiler and the compiler flags, and
\emph{(vi)}
the number of threads started.


\subsection*{Test Set of Initial Value Problems}
\label{sec:testbed:ivp}


In our experiments, we consider a broad set of IVPs (Table \ref{tab:ivps}) that 
exhibit different characteristics:
\emph{(i)} \emph{Cusp} combines Zeeman's cusp catastrophe model for a 
threshold-nerve-impulse mechanism with the van der Pol oscillator 
\cite{Hairer2002},
\emph{(ii)} \emph{IC} describes a traversing signal through a chain of 
$N$ concatenated inverters \cite{Bartel2002},
\emph{(iii)} \emph{Medakzo} describes the penetration of radio-labeled 
antibodies into a tissue infected by a tumor \cite{Mazzia2008}, and
\emph{(iv)} \emph{Wave1D} describes the propagation of disturbances at a fixed 
speed in one direction \cite{Calvo2004}.


\begin{table}[tbp]
  \centering
  \caption{Characteristics of the test set of IVPs.}
  \label{tab:ivps}
  \begin{threeparttable}{}
  \begin{tabular}{lcccc}
    \toprule
    \bfseries{IVP} & Cusp & IC & Medakzo & Wave1D\\
    \midrule
    \bfseries{Acces distance}\tnote{1}
    & unlimited & limited & unlimited & limited\\    
    \bfseries{Computational behavior} & mixed & compute-bound & 
    mixed & memory-bound\\
    \bottomrule
  \end{tabular}
  \begin{tablenotes}
    \item[1] In practice, many IVPs are sparse, i.e. only access few components 
    of $\Y$ when evaluating function $\f$ (line 4, Lst. \ref{lst:pirk}). A 
    special case of sparse is \emph{limited access distance} $d(\f)$, where 
    $f_j$ only accesses components $y_{j-d(\f)}$ to $y_{j+d(\f)}$.
  \end{tablenotes}
  \end{threeparttable}  
\end{table}


\subsection*{Test Set of Target Platforms}
\label{sec:testbed:machine}


We conducted our experiments on three different shared-memory systems (Table 
\ref{tab:platforms}). For all experiments, the CPU clock was fixed, 
hyper-threading disabled and thread binding set with 
\emph{KMP\_AFFINITY=granularity=fine,compact}. All codes were compiled with the 
Intel C compiler and compiler flags \emph{-O3}, \emph{-xAVX} and 
\emph{-fno-alias} set.













\begin{table*}[tbp]
  \centering
  \caption{Characteristics of the test set of target platforms.}
  \label{tab:platforms}
  \begin{threeparttable}{}
  \begin{tabular}{lccc}
    \toprule
    \bfseries{Name} & HSW & IVB & SKY\\
    \midrule
    \bfseries{Micro-architecture} & Haswell EP & Ivy-Bridge EP & Skylake SP \\
    \bfseries{CPU} & Xeon E5-2630 v3 & Xeon E5-2660 v2 & Xeon Gold 6148\\
    \bfseries{Clock speed} & 2.3 GHz & 2.2 GHz & 1.76 GHz\\
    \bfseries{Cores} & 8 & 10 & 20\\
    \midrule
    \bfseries{L1 cache} (data) & 32 kB & 32 kB & 32 kB\\
    \bfseries{L2 cache} & 256 kB & 256 kB & 1 MB\\
    \bfseries{L3 cache} (shared) & 20 MB & 25 MB & 27.5 MB\\
    \bfseries{Cache line size} & 64 B & 64 B & 64 B\\
    \midrule
    \midrule
    \multicolumn{4}{l}{\bfseries{Instruction throughput per cycle} (double precision)}\\
    ADD / FMA / MUL & 2 / 1 / 1 & 1 / - / 1 & 4 / 2 / 4\\
    \midrule
    \midrule
    \bfseries{Compiler} & icc 19.0.5 & icc 19.0.4 & icc 19.0.2\\
    \bottomrule
  \end{tabular}
  \end{threeparttable}
\end{table*}


\section{Offsite Autotuning Approach}
\label{sec:offsite}


In this work, we introduce the \highlight{Offsite} \offline~AT approach on the 
example of explicit ODE methods. Before starting a new Offsite run, the 
\emph{tuning scenario} desired, which consists of: \emph{(i)} the pool of 
possible implementations and program transformations, \emph{(ii)} the ODE base 
method(s), \emph{(iii)} the IVP(s), and \emph{(iv)} the target platform, is 
defined using description languages in the YAML standard\footnote{YAML is a data
serialization language; \url{https://yaml.org}}.

\begin{figure}[tbp]
  \centering
  \begin{tikzpicture}[%
  node distance=11em,
  block/.style={rectangle, draw=green, thick, fill=green!25, text width=6.0em, align=center, minimum size=0.2em, font=\tiny},
  input/.style={rectangle, draw=yellow, thick, fill=yellow!25, text width=6.0em, align=center, rounded corners, minimum size=1.25em, font=\tiny},
  line/.style={draw,->,thick, shorten >=2pt},  
  frame/.style={rectangle, draw=black, thick, fill=gray!5, rounded corners}, 
  database/.style={draw, cylinder, cylinder uses custom fill, cylinder body fill=gray!65, cylinder end fill=gray!50, shape border rotate=90, aspect=0.2, draw=gray, minimum size=1.125em, font=\tiny}]
  \draw[frame] (0.7,1.45) rectangle (11.75,-1.45);
  
  \draw (0.0,0.0) node[input, rotate=90, text width=5.225em] (node_input) {Tuning Scenario};
  \draw (2.25,1.05) node[block] (node_kernel_workingsets) {Kernel\\ Working Sets};
  \draw (2.25,0.35) node[block] (node_kernel_generator) {Kernel Code Generation};  
  \draw (2.25,-0.35) node[block] (node_impl_variants) {Derived Impl. Variants};
  \draw (2.25,-1.05) node[block] (node_communication_costs) {Communication Cost Benchmarks};
  \draw[line, dashed] (node_input.south) -- (node_kernel_generator.west);
  \draw[line, dashed] (node_input.east) |- (node_kernel_workingsets.west);
  \draw[line, dashed] (node_input.south) -- (node_impl_variants.west);
  \draw[line, dashed] (node_input.west) |- (node_communication_costs.west);
  
  \draw (5.0,0.65) node[block] (node_kernel_prediction) {Kernel prediction};
  \draw[line] (node_kernel_generator.east) -- (node_kernel_prediction.west);
  \draw[line] (node_kernel_workingsets.east) -| (node_kernel_prediction.north);
  
  \draw (5.0,-0.35) node[block] (node_impl_prediction) {Impl. Variant Prediction};
  \draw[line] (node_kernel_prediction.south) -- (node_impl_prediction.north);
  \draw[line] (node_impl_variants.east) -- (node_impl_prediction.west);
  \draw[line] (node_communication_costs.east) -| (node_impl_prediction.south);  
  
  \draw (7.75,-0.35) node[block] (node_ranking) {Impl. Variant\\ Ranking};
  \draw[line] (node_impl_prediction.east) --(node_ranking.west);
    
  \draw (10.5,-0.35) node[block] (node_impl_generator) {Impl. Variant Code Generation};
  \draw[line] (node_ranking.east) -- (node_impl_generator.west);
  
  \draw (10.5,-1.85) node[database] (node_db) {\textbf{DB}};
  \draw (11.25,-1.45) node (node_frame_border) {};
  \draw[line, dashed, <->] (node_frame_border.east) |- (node_db.east);
  
  \draw (8.5, -1.85) node[input, draw=blue, fill=blue!25] (node_online_tuning) {Online AT};
  \draw[line, dashed, <->] (node_online_tuning.east) -- (node_db.west);
  
  \draw (11.0,1.1) node[thick, align=center, text=gray] (label_offsite) {\small\textbf{\textit{Offsite}}};
\end{tikzpicture}
  \caption{Workflow of the Offsite autotuning approach.}
  \label{fig:offsite_workflow}
\end{figure}
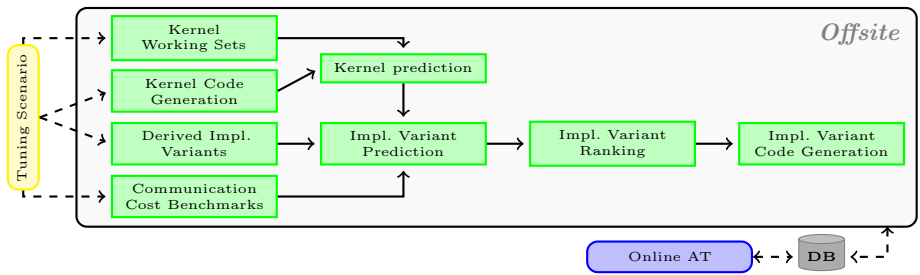

From its input data, Offsite automatically handles the whole tuning workflow 
(Fig. \ref{fig:offsite_workflow}). First, Offsite generates optimized,
platform-specific and problem-specific code for all kernels and derives all 
possible implementation variants. Applying an analytic performance prediction
methodology, the performance of each kernel is predicted for either \emph{(i)} a
fixed ODE system size $n$---if specified by the user or prescribed by the 
ODE\footnote{There are scalable ODE systems but also ODEs with a fixed size 
\cite{Hairer2002}.}---or \emph{(ii)} a set of relevant ODE system sizes 
determined by a working set model. The performance of a variant is derived by 
combining the predictions of its kernels and adding an estimate of its 
synchronization costs. Variants are ranked by their performance to identify the 
most efficient variant(s). All obtained prediction and ranking data are stored 
in a database. For the best ranked variants, Offsite generates optimized, 
platform-specific and problem-specific code.


\subsection{Input Description Languages}
\label{sec:dsl}


A decisive, yet cumbersome step in AT is generating optimized code. Often, there
is a large pool of possible implementation variants, applicable program 
transformations (e.g. loop transformations) and tunable parameters (e.g. tile 
sizes) available. Furthermore, exploiting characteristics of the input data can 
enable more optimizations (e.g. constant propagation). Writing all variants by 
hand, however, would be tedious and error-prone and there is demand for 
automation.
In this work, we introduce multilevel description languages to describe 
implementations, ODE methods, IVPs and target platforms in an abstract way. 
Offsite can interpret these languages and automatically derives optimized code.


\subsubsection{The Base ODE Method}
\label{sec:dsl:method}


of a PIRK method is characterized by its Butcher table---i.e., coefficient 
matrix $A$, weight vector $b$, node vector $c$---and a small set of properties: 
\emph{(i)} number of stages $s$, \emph{(ii)} order $o$, \emph{(iii)} number of 
corrector steps $m$. Exploiting these properties, however, can have a large 
impact on the efficiency of an implementation variant and should be included 
into the code generation in order to obtain the most efficient code. The 
$i$-loop in Listing \ref{lst:yaml_kernel_approx}, e.g., might be replaceable by 
a single vector operation for specific $s$, or zero entries in the Butcher table
might allow to save computations.

Listing \ref{lst:yaml_ode} shows the ODE method description format on the 
example of \emph{Radau II\,A(7)} which is a four-stage method with order seven  
applying six corrector steps per timestep. To save space, only an excerpt of 
the Butcher table is shown with a reduced number of digits.


\subsubsection{IVPs}
\label{sec:dsl:ivp}


are described in the IVP description format shown by \emph{IC} (Lst. 
\ref{lst:yaml_ivp}):

\appar{(i)}
\dslKey{components} describes the $n$ components of the IVP. Each component 
contains a \dslKey{code} YAML block that describes how function evaluation 
$\f(\tk + c_i \hk, \Y_i^{(k-1)})$ (l. 4, Lst. \ref{lst:pirk}) will be 
substituted during code generation whereby \dslKey{\%in} is a placeholder for 
the used input vector $\Y_i^{(k-1)}$. Adjacent components that execute the same 
computation can be described by a single block whereby \dslKey{first} denotes 
the first component and \dslKey{size} specifies the total number of adjacent 
components handled by that particular block.

\appar{(ii)}
\dslKey{constants} defines IVP-specific parameters replaced with their actual 
values during code generation and might possibly enable further code 
optimizations. In IVP \emph{IC}, e.g., a multiplication could be saved if the 
given electrical resistance \emph{R} equals $1.0$.


\begin{figure}[tbp]
\begin{minipage}[t]{0.4225\textwidth}
  \centering
  \input{fig/listings/yaml_ode.tex}
\end{minipage}
\hfill
\begin{minipage}[t]{0.5\textwidth}
  \centering
  \input{fig/listings/yaml_ivp.tex}
\end{minipage}
\vspace{-3ex}
\end{figure}


\subsubsection{Target Platform and Compiler}
\label{sec:dsl:machine}


are described using the machine description format introduced by 
Kerncraft\footnote{For example files, we refer to 
\url{https://github.com/RRZE-HPC/kerncraft}.}. Its general structure is 
tripartite: \emph{(i)} the execution architecture description, \emph{(ii)} the 
cache and memory hierarchy description, and \emph{(iii)} benchmark results of 
typical streaming kernels.


\subsubsection{Implementation Variants}
\label{sec:dsl:implementations}


\begin{figure}[tbp]
\begin{minipage}[t]{0.465\textwidth}
  \centering
  \input{fig/listings/yaml_template.tex}
  \input{fig/listings/code_kerncraft_kernel_approx.tex}
\end{minipage}
\hfill
\begin{minipage}[t]{0.4625\textwidth}
  \centering
  \input{fig/listings/yaml_skeleton.tex}
  \input{fig/listings/code_impl_kernel_approx.tex}
\end{minipage}
\vspace{-3ex}
\end{figure}

of numerical algorithms are abstracted by description languages as 
(i)~\highlight{kernel templates} and (ii)~\highlight{implementation skeletons}.

\emph{Kernel Templates} define basic computation kernels and possible variations
of this kernel enabled by program transformations that preserve semantic 
correctness. Listing \ref{lst:yaml_kernel_approx} shows the kernel template 
description format on the example of \emph{APRX}, which covers computation 
$\sum_{i=1}^{s}b_i\F_i^{(m)}$ (l. 5, Lst. \ref{lst:pirk}):

\appar{(i)} 
\dslKey{datastructs} defines required data structures.

\appar{(ii)}
\dslKey{computations} describes the computations covered by a kernel template. 
Each computation corresponds to a single line of code and has an unique 
identifier (E.g. \emph{C1} in Lst. \ref{lst:yaml_kernel_approx}). Computations 
can contain IVP evaluations which are marked by keyword \dslKey{\%RHS} and are 
replaced by an IVP component during code generation (E.g. for \emph{IC} by
line 5 of Lst. \ref{lst:yaml_ivp}). Hence, if a kernel template contains 
\dslKey{\%RHS}, a separate, specialized kernel version has to be generated for 
each IVP component.

\appar{(iii)}
\dslKey{variants} contains possible kernels of a kernel template enabled by 
program transformations. For each kernel, its workings sets (\dslKey{working 
sets}) and its program code (\dslKey{code}) are specified. A \dslKey{code} block
defines for a kernel its order of computations and program transformations and 
using four keywords. Computations are specified by keyword \dslKey{\%COMP} whose
parameter must correspond to one of the identifiers defined in the 
\dslKey{computations} block (E.g. \emph{C1} in Lst. 
\ref{lst:yaml_kernel_approx}). For-loop statements are defined by 
\dslKey{\%LOOP\_START} and \dslKey{\%LOOP\_END}. The first parameter of 
\dslKey{\%LOOP\_START} specifies the loop variable name, the second the number 
of loop iterations and an optional third parameter \dslKey{unroll} indicates 
that the loop will be unrolled during code generation. Loop-specific pragmas can
be added using \dslKey{\%PRAGMA}.

\emph{Implementation skeletons} define processing orders of kernel templates and
required communication points. From skeletons, concrete implementation variants 
are derived by replacing its templates with concrete kernel code. Listing 
\ref{lst:yaml_impl_A} shows the implementation skeleton description format on 
the example of skeleton \emph{A} which is a realization of a PIRK method (Lst. 
\ref{lst:pirk}) that focuses on parallelism across the ODE system, i.e its $n$ 
equations are distributed blockwise among the threads. \emph{A} contains a loop 
$k$ over the $m$ corrector steps dividing each corrector step into two 
templates: \emph{RHS} computes the IVP function evaluations (l. 5, 
Lst.~\ref{lst:pirk}) which are used to compute the linear combinations (l. 4, 
Lst. \ref{lst:pirk}) in \emph{LC}. Per corrector step, two synchronizations are 
needed as \emph{RHS}---depending on the IVP solved---can potentially require all
components of the linear combinations from the last iteration of $k$. After all 
corrector steps are computed, the next approximation $\y_{\kappa + 1}$ is 
calculated by templates \emph{APRX} and \emph{UPD} (l. 6, Lst. \ref{lst:pirk}).
Four keywords suffice to specify skeletons:

\appar{(i)} \dslKey{\%LOOP\_START} and \dslKey{\%LOOP\_END} define for-loops.

\appar{(ii)} \dslKey{\%COM} states communication operations of an implementation
skeleton. Skeleton \emph{A}, e.g., requires $2m + 2$ barrier synchronizations.

\appar{(iii)} \dslKey{\%KERNEL} specifies an executed kernel template. Its 
parameter must correspond to the name of a kernel template. During code 
generation \dslKey{\%KERNEL} is replaced by actual kernel code (e.g. \emph{APRX}
in Lst. \ref{lst:code_impl_kernel_approx_ji}).




\subsection{Rating Implementation Variant Performance}
\label{sec:offsite:rating_variants}


Offsite can automatically identify the most efficient implementation variant(s) 
from a pool of available variants using analytic performance modelling (Fig. 
\ref{fig:offsite_workflow}):

\appar{(i)}
In a first step, Offsite automatically generates code for all kernels in a 
special code format processable by kerncraft\footnote{In this work, version 
\emph{0.8.3} of the Kerncraft tool was used.}. Kernel code generation 
(\emph{Kernel Code Generation} in Fig. \ref{fig:offsite_workflow}) includes 
specializations of the code on the target platform, IVP, ODE method and (if 
fixed) ODE system size $n$. Listing \ref{lst:code_kerncraft_kernel_approx_ji} 
exemplary shows the code generated for kernel \emph{APRX\_ji} of kernel template
\emph{APRX} (Lst. \ref{lst:yaml_kernel_approx}) when specialized on ODE method 
\emph{Radau II\,A(7)} and $n=161$. As specified in the template description, $j$
loop is unrolled completely. Further, Butcher table coefficients ($\vect{b}$) 
and known constants ($s=4$, $n=161$) are substituted.


\appar{(ii)}
In some tuning scenarios, the ODE system size $n$ is not yet known during 
installation time. Giving predictions for all valid $n$ values, however, is in 
general not feasible. By applying a working set model (Sec. 
\ref{sec:offsite:working_set_model}), Offsite automatically determines for each 
kernel a set of relevant $n$ (\emph{Kernel Working Sets}, Fig. 
\ref{fig:offsite_workflow}) for which predictions are then obtained in the next 
step.


\appar{(iii)}
Offsite automatically computes node-level runtime predictions (Sec. 
\ref{sec:offsite:performance_methodology}) for each implementation variant 
(\emph{Impl. Variant Prediction}, Fig. \ref{fig:offsite_workflow}) by adding up 
the kernel predictions of its kernels and adding an estimate of its 
communication costs (\emph{Communication Cost Benchmarks}, Fig. 
\ref{fig:offsite_workflow}), which Offsite derives from benchmark data. For each
of the kernel codes generated in step \emph{(i)}, its kernel prediction is 
automatically derived by Offsite (\emph{Kernel Prediction}, Fig. 
\ref{fig:offsite_workflow}) whereby Kerncraft is used to construct the ECM 
model.


\appar{(iv)}
Using these node-level runtime predictions, Offsite ranks implementation 
variants by their performance (\emph{Impl. Variant Ranking}, Fig. 
\ref{fig:offsite_workflow}).


\appar{(v)}
From the ranking of implementation variants, Offsite automatically derives the 
subset $\Lambda$ of the best rated variant(s) which contains all variants 
$\lambda$ whose performance is within an user-provided maximum deviation from 
the best rated variant. For each variant of $\lambda$, Offsite generates 
optimized, platform-specific and problem-specific code (\emph{Impl. Variant Code
Generation}, Fig. \ref{fig:offsite_workflow}).
Listing \ref{lst:code_impl_kernel_approx_ji} shows an excerpt of the code 
generated for an variant of implementation skeleton \emph{A} which substitutes 
kernel template \emph{APRX} with kernel \emph{APRX\_ji} and was specialized on 
ODE method \emph{Radau II\,A(7)}, IVP \emph{IC} and $n=161$.


\subsection{Performance Prediction Methodology}
\label{sec:offsite:performance_methodology}


The performance prediction methodology applied by Offsite expands on the works 
of \cite{Seiferth2018} and comprises: \emph{(i)} a node-level runtime prediction
of an implementation variant and \emph{(ii)} an estimate of its intra-node 
communication costs.


\subsubsection{Node-level Runtime Prediction.}
\label{sec:offsite:performance_methodology:node_level}


Base of the node-level prediction is the analytic \emph{ECM 
(Execution-Cache-Memory) performance model}. For an in-depth explanation, we refer to~\cite{Stengel2015,Hofmann2019}. The ECM model gives an 
estimation of the number of CPU cycles per cache line (CL) required to execute a
particular loop kernel on a multi- or many-core chip which includes 
contributions from the in-core execution time $T_\textrm{core}$ and the data 
transfer time $T_\textrm{data}$:
\begin{align}
T_\textrm{core} & = \textrm{max}(T_\textrm{OL},\enspace T_\textrm{nOL})\enspace, \label{eq:ecm:core}\\
T_\textrm{data}^{\textrm{L3}} & = T_\textrm{L1L2}^{\textrm{data}} + T_\textrm{L2L3}^{\textrm{data}} + T_{\textrm{L2L3}}^{\textrm{p}}\enspace. \label{eq:ecm:data_L3}
\end{align}
$T_\textrm{core}$ is defined as the time required to retire the instructions of 
a single loop iteration under the assumptions that (i)~there are no loop-carried
dependencies, (ii)~all data are in the L1 data cache, (iii)~all instructions are
scheduled independently to the ports of the units, and (iv)~the time to retire 
arithmetic instructions and load/store operations can overlap due to speculative
execution depending on the target platform. Hence, the unit that takes the 
longest to retire its instructions determines $T_\textrm{core}$.
$T_\textrm{data}^{\textrm{level}}$ factors in the time required to transfer all 
data from its current location in the memory hierarchy to the L1 data cache and 
back. The single contributions of transfers between levels $i$ and $j$ of the 
memory hierarchy $T_{ij}^{\textrm{data}}$ are determined depending on the
amount of transferred CLs. Depending on the platform used, an optional latency 
penalty $T_{ij}^p$ might be added. In~\eqref{eq:ecm:data_L3} 
$T_\textrm{data}^{\textrm{level}}$ is exemplarily shown for data coming from the
L3 cache under the assumption that a latency penalty between L2 and L3 cache has
to be factored in on the platform used.
Combining all contributions, a single-core prediction
\begin{equation}\label{eq:ecm:prediction}
T_\textrm{ECM}^{\textrm{level}} = \textrm{max}(T_\textrm{OL},\enspace T_\textrm{nOL} + T_\textrm{data}^{\textrm{level}})
\end{equation}
can be derived, whereby the overlapping capabilities of the target platform 
determine whether a contribution is considered overlapping or non-overlapping.

Using Kerncraft, Offsite obtains ECM model predictions \eqref{eq:ecm:prediction}
for all kernels $\lambda$. For each kernel, \emph{kernel runtime prediction}
\begin{equation}\label{eq:node_level:prediction:kernel_time}
\phi_{\lambda} = \frac{\alpha_{\lambda} \cdot \beta_{\lambda}}{\delta \cdot f}
\end{equation}
yields the runtime in seconds of kernel $\lambda$, where $\alpha_\lambda$ is 
\eqref{eq:ecm:prediction} computed for a specific number of running cores 
$\tau$, $\beta_\lambda$ is the number of loop iterations executed, $\delta$ is 
the number of data elements fitting into one CL and $f$ is the CPU frequency.
By summing up the individual kernel runtime predictions $\phi_{\lambda}$ of its 
basic kernels $\lambda$ and adding an estimate of its communication costs 
$t_\textrm{com}$, the \emph{node-level runtime prediction} $\theta_{\epsilon}$ 
of an implementation variant $\epsilon$ is given by:
\begin{equation}\label{eq:node_level:prediction:node_level_runtime}
\theta_{\epsilon} = \sum^{}_{\lambda}{\phi_\lambda} + t_{\textrm{com}}\enspace.
\end{equation}

\noindent
\emph{Remark:}
\cite{Seiferth2018} used an older \emph{Kerncraft} version that could not yet 
return ECM predictions for multiple core counts $\tau$ with a single run, but 
additionally returned the kernel's saturation point  $\sigma_\lambda$. Hence, an
extra factor $\min(\tau, \sigma_\lambda)$ was needed in the denominator of 
\eqref{eq:node_level:prediction:kernel_time} in their work.


\subsubsection{Estimate of Intra-node Communication Costs.}
\label{sec:offsite:performance_methodology:communication}


For the implementation variants considered in this work, the costs of the 
required OpenMP barrier operations are estimated depending on the number of 
threads using linear regression.


\subsubsection{Reusability of Performance Predictions.}
\label{sec:offsite:performance_methodology:re_use}


Prediction data (e.g. kernel runtime predictions) collected for a specific 
implementation variant can be reused to estimate other variants (if they also 
include that kernel) or to estimate other IVPs (if the kernel does not contain 
any IVP evaluations). In the context of AT, this is a decisive advantage 
compared to time-consuming runtime testing of variants which requires running 
each additionally added variant as well or to run all variants anew when 
changing the IVP solved.


\subsection{Working Set Model}
\label{sec:offsite:working_set_model}


If the ODE system size $n$ is not fixed---either by the user or restrictions of 
the IVP---selecting the most efficient implementation variant(s) at 
installation time leads to an exhaustive search over the possibly vast space of 
values for $n$. To minimize the number of predictions required per kernel, the 
set of estimated $n$ values is reduced by a model-based restriction, 
the \highlight{working set} of the kernel, which corresponds to the amount of
data referenced by a kernel.

We use the working sets
to identify for each kernel the maximum $n$ that still fit into the single cache
levels. Using these maximums, ranges of consecutive $n$ values for which the ECM
prediction \eqref{eq:ecm:prediction} stays constant\footnote{The ECM prediction 
factors in the location of data in the memory hierarchy. As a simplified 
assumption---neglecting overlapping effects at cache borders---, this means that
as long as data locations do not change, the ECM model yields the same value for
a kernel independent from the actual $n$.} can be derived. The medium values of 
these ranges form the working set of the kernel.


\section{Experimental Evaluation}
\label{sec:experiments}


We validate Offsite using the experimental test bed introduced in Section 
\ref{sec:testbed}. In particular, we study the efficiency of Offsite in four AT 
scenarios when tuning four different IVPs on three different target platforms 
and compare the efficiency of the following AT strategies:
\begin{itemize}
\itemLabel{(i)}
\emph{BestVariant} covers the case that the most efficient implementation
variant is already known (e.g. from previous execution) and no AT is required. 
\itemLabel{(ii)}
\emph{RunAll} runs all variants in order to identify the most efficient 
variant.
\itemLabel{(iii)}
\emph{OffsitePreselect5} runs an Offsite determined subset of all 
variants, which contains all variants withing a 5 \% deviation of the best 
ranked variant, to identify the most efficient variant of that subset.
\itemLabel{(iv)}
\emph{OffsitePreselect10} runs variants within a 10 \% deviation of the best 
variant.
\itemLabel{(v)}
\emph{RandomSelect} randomly runs 20 of the total 56 variants.
\end{itemize}


\subsection{Derived Implementation Variants}
\label{sec:experiments:impl_variants}


Table \ref{tab:impl_variants} summarizes the implementation skeletons and kernel
templates used in this work. In total, we consider eight skeletons from which 56
implementation variants can be derived. Each table column shows the templates 
required by a particular skeleton. E.g., skeleton \emph{A} (Lst. 
\ref{lst:yaml_impl_A}) uses templates \emph{LC}, \emph{RHS}, \emph{APRX} and 
\emph{UPD} and twelve variants can be derived from \emph{A} as there are six 
different kernels of \emph{LC} (enabled by loop interchanges, unrolls, pragmas) 
and two of \emph{APRX}.

In total, 17 different kernels can be derived from the eight kernel templates 
available. To predict the performance of all 56 variants, only these 17 kernels 
have to be estimated. Further, when obtaining predictions off all 56 variants 
for a different IVP, only those four templates that contain IVP 
evaluations---and thus their six corresponding kernels---need to be 
re-evaluated, while prediction data of the remaining kernels can be retrieved 
from database.


\begin{table*}[tbp]
  \centering
  \caption{Overview of the implementation variants considered.}
  \label{tab:impl_variants}
  \resizebox{\textwidth}{!}{%
  \begin{threeparttable}
  \begin{tabular}{cc|cccccccc}
    \toprule
    \multicolumn{2}{c|}{\bfseries{Impl. Skeleton}}
    & \multicolumn{8}{c}{\bfseries{Kernel Template}\tnote{1} (\#Kernels)} \\
    \multicolumn{2}{c|}{(\#Impl. Variants)}
    & LC(6) & RHS*(1) & RHSLC*(1) & APRX(2)
    & RHSAPRX*(2) & UPD(1) & APRXUPD(2) & RHSAPRXUPD*(2) \\
    \midrule
    A & (12) & x & x &   & x &   & x &   &   \\
    B & (12) & x & x &   &   &   &   & x &   \\
    C & (2)  &   &   & x & x &   & x &   &   \\
    D & (2)  &   &   & x &   &   &   & x &   \\
    E & (2)  &   &   & x &   & x & x &   &   \\
    F & (2)  &   &   & x &   &   &   &   & x \\
    G & (12) & x & x &   &   & x & x &   &   \\
    H & (12) & x & x &   &   &   &   &   & x\\
    \bottomrule
  \end{tabular}
  \begin{tablenotes}
    \item[1] A kernel template marked with \textbf{*} contains evaluations of 
    the IVP.
  \end{tablenotes}  
  \end{threeparttable}
  }
\end{table*}


\subsection{AT Scenario -- All Input Known}
\label{sec:experiments:fixed_n}


As first test scenario, we consider the case that all input is known at 
installation time, in particular the ODE system size $n$. In such cases, Offsite
is applied without the working set model. Performance predictions, however, are 
only obtained for that particular $n$ and a new Offsite AT run would be required
if $n$ changes.

Table \ref{tab:at_fixed_n} compares the accuracy and efficiency of AT strategies
when tuning four different IVPs on three different target platforms for $n = 
36{,}000{,}000$ and ODE method \emph{Radau II\,A(7)}. For AT strategies
\emph{OffsitePreselect5} and \emph{OffsitePreselect10}, $t_\textrm{step}$ yields
the time in seconds it takes to execute a timestep using the measured best 
implementation variant from the subset $\Lambda$ of variants $\lambda$ tested by
that strategy. \emph{Performance loss} denotes the percent runtime deviation of 
that particular measured best variant from the variant selected by 
\emph{BestVariant} ($t_\textrm{best}$). Ideally, an AT strategy correctly 
identifies the measured best variant and, thus, would suffer no performance 
loss. For an AT strategy, \emph{$\vert \Lambda \vert$} yields the cardinality of
subset $\Lambda$ and the percent \emph{tuning overhead} of applying that 
strategy is defined as $\frac{t_\textrm{tune} - \vert \Lambda \vert 
t_{\textrm{best}}}{\vert \Lambda \vert t_{\textrm{best}}} \cdot 100$ where 
$t_\textrm{tune} = \sum_\lambda^\Lambda t_\lambda$ is the time required to test 
all variants and $\vert \Lambda \vert t_{\textrm{best}}$ is the time needed to 
execute the measured best variant instead.


\subsubsection{Haswell.}
\label{sec:experiments:fixed_n:haswell}


AT strategy \emph{RunAll} causes a significant tuning overhead for all 
IVPs, while \emph{OffsitePreselect5} and \emph{OffsitePreselect10} only lead to 
marginal overhead as the subset of tested variants is considerably smaller, 
while still being able to select the measured best variant for all IVPs but 
\emph{Wave1D}.


\subsubsection{IvyBridge.}
\label{sec:experiments:fixed_n:ivy}


Again, \emph{RunAll} leads to decisive overhead compared to both Offsite 
strategies and the measured best variant is correctly identified for all IVPs. 
However, for IVP \emph{IC} only \emph{OffsitePreselect10} finds the best 
variant. As \emph{IC} is compute-bound (Table \ref{tab:ivps}), the IVP 
evaluation dominates the computation time while the order of the remaining 
computations has only minor impact. Hence, already minor jitter can lead to a 
different variant being selected.


\subsubsection{Skylake.}
\label{sec:experiments:fixed_n:skylake}


Similar observations as on the two previous systems can be made. The overhead of
both Offsite strategies is marginal compared to \emph{RunAll}. For all IVPs, the
measured best variant is successfully identified.


\begin{table*}[t]
  \setlength{\tabcolsep}{3.5pt}
  \centering
  \caption{Comparison of different AT strategies applied to four different IVPs 
  with $n = 36{,}000{,}000$ and \emph{Radau II\,A(7)}.}
  \label{tab:at_fixed_n}
  \resizebox{\textwidth}{!}{  
  \begin{tabular}{c|l|rrrr}
    \toprule
    & \bfseries{IVP}
    & \multicolumn{1}{c}{\bfseries{Cusp}}
    & \multicolumn{1}{c}{\bfseries{IC}}
    & \multicolumn{1}{c}{\bfseries{Medakzo}}
    & \multicolumn{1}{c}{\bfseries{Wave1D}}\\
    
    \midrule
    
    \parbox[t]{3mm}{\multirow{10}{*}{\rotatebox{90}{\bfseries{Haswell (8 cores)}}}}
    & \bfseries{BestVariant}
    & {\footnotesize{F\_*ji}} & {\footnotesize{F\_*ij}} & {\footnotesize{F\_*ji}} & {\footnotesize{H\_LCjli\_*ij}} \\
    & \bfseries{BestVariant} {$t_{\textrm{step}}$[s]}
    & 1.28 & 0.80 & 1.29 & 1.04 \\
    
    & \multicolumn{5}{l}{\bfseries{AT strategy -- OffsitePreselect5}} \\
    & $\vert \Lambda \vert$ (tuning overhead)
    & 3 (1\%) & 3 (3\%) & 2 (2\%) & 3 (5\%) \\    
    & $t_{\textrm{step}}$[s] selected variant (perf. loss)
    & 1.28 (--) & 0.80 (--) & 1.29 (--) & 1.08 (4\%) \\
    
    & \multicolumn{5}{l}{\bfseries{AT strategy -- OffsitePreselect10}} \\
    & $\vert \Lambda \vert$ (tuning overhead)
    & 3 (1\%) & 3 (3\%) & 3 (1\%) & 4 (5\%) \\    
    & $t_{\textrm{step}}$[s] selected variant (perf. loss)
    & 1.28 (--) & 0.80 (--) & 1.29 (--) & 1.08 (4\%) \\
    
    & \multicolumn{5}{l}{\bfseries{AT strategy -- RunAll}} \\
    & $\vert \Lambda \vert$ (tuning overhead)
    & 56 (42\%) & 56 (44\%) & 56 (20\%) & 56 (16\%) \\
    
    \midrule

    \parbox[t]{3mm}{\multirow{10}{*}{\rotatebox{90}{\bfseries{IvyBridge (10 cores)}}}}
    & \bfseries{BestVariant}
    & {\footnotesize{E\_*ji}} & {\footnotesize{F\_*ij}} & {\footnotesize{F\_*ji}} & {\footnotesize{F\_*ji}} \\    
    & \bfseries{BestVariant} $t_{\textrm{step}}$[s]
    & 1.16 & 0.725 & 1.20 & 1.04\\  
    
    & \multicolumn{5}{l}{\bfseries{AT strategy -- OffsitePreselect5}} \\
    & $\vert \Lambda \vert$ (tuning overhead)
    & 3 (3\%) & 1 (1\%) & 3 (1\%) & 3 (0.4\%) \\ 
    & $t_{\textrm{step}}$[s] selected variant (perf. loss)
    & 1.16 (--) & 0.734 (1\%) & 1.20 (--) & 1.04 (--) \\

    & \multicolumn{5}{l}{\bfseries{AT strategy -- OffsitePreselect10}} \\
    & $\vert \Lambda \vert$ (tuning overhead)
    & 3 (3\%) & 3 (3\%) & 3 (1\%) & 3 (0.4\%) \\
    & $t_{\textrm{step}}$[s] selected variant (perf. loss)
    & 1.16 (--) & 0.725 (--) & 1.20 (--) & 1.04 (--) \\
    
    & \multicolumn{5}{l}{\bfseries{AT strategy -- RunAll}} \\
    & $\vert \Lambda \vert$ (tuning overhead)
    & 56 (54\%) & 56 (59\%) & 56 (41\%) & 56 (44\%) \\
    
    \midrule
    
    \parbox[t]{3mm}{\multirow{10}{*}{\rotatebox{90}{\bfseries{Skylake (20 cores)}}}}
    & \bfseries{BestVariant}    
    & {\footnotesize{E\_*ji}} & {\footnotesize{F\_*ji}} & {\footnotesize{F\_*ji}} & {\footnotesize{F\_*ji}} \\     
    & \bfseries{BestVariant} $t_{\textrm{step}}$[s]
    & 0.43 & 0.24 & 0.45 & 0.40\\ 

    & \multicolumn{5}{l}{\bfseries{AT strategy -- OffsitePreselect5}} \\
    & $\vert \Lambda \vert$ (tuning overhead)
    & 3 (1\%) & 3 (2\%) & 3 (1\%) & 3 (\%) \\
    & $t_{\textrm{step}}$[s] selected variant (perf. loss)
    & 0.43 (--) & 0.24 (--) & 0.45 (--) & 0.40 (--) \\
    
    & \multicolumn{5}{l}{\bfseries{AT strategy -- OffsitePreselect10}} \\
    & $\vert \Lambda \vert$ (tuning overhead)
    & 3 (1\%) & 3 (2\%) & 3 (1\%) & 3 (1\%) \\
    & $t_{\textrm{step}}$[s] selected variant (perf. loss)
    & 0.43 (--) & 0.24 (--) & 0.45 (--) & 0.40 (--) \\    
    
    & \multicolumn{5}{l}{\bfseries{AT strategy -- RunAll}} \\
    & $\vert \Lambda \vert$ (tuning overhead)
    & 56 (69\%) & 56 (65\%) & 56 (41\%) & 56 (44\%) \\
    \bottomrule    
  \end{tabular}
  }
\end{table*}


\subsection{AT Scenario -- Unknown ODE System Size}
\label{sec:experiments:unknown_n}


The next scenario considered is that of a still unknown ODE system size $n$ at 
installation time. In these cases, the working set model is applied to determine
a set of sample $n$ values for which Offsite computes predictions and from which
predictions for the whole range of possible $n$ are derived. As this requires 
computing multiple performance predictions, a single Offsite run takes longer 
than in the previous scenario. This particular Offsite run, however, already 
covers all possible $n$ and no further run will be required when switching $n$
at a later point.

Figures \ref{fig:at_cusp_var} and \ref{fig:at_ic_var} show for the single 
implementation variants selected as best variant by the AT strategies 
considered, the time per timestep of \emph{IC} and \emph{Cusp} on three 
platforms (each using their max. number of cores). On the x-axis, $n$ is plotted
up to $n=60{,}000{,}000$. The y-axis shows the time per component of $n$ in 
seconds needed by a specific variant to solve a timestep for \emph{Radau 
II\,A(7)}.

\begin{figure*}[tbp]
  \centering
  \hfill
  \scalebox{.93}{\begin{tikzpicture}[y=.2cm, x=.7cm]
  \centering
  \begin{axis}[
    height=4.5cm,
    width=0.4\textwidth,
    grid=major,
    tick align=inside,
    xlabel={n},
    ylabel={Time per component [s/n]},
    xmin=0,
    xmax=3.6e+07,
    ymin=3.4e-08,
    ymax=4.2e-08,
  ]
  \addplot[red, mark=*, mark options={fill=white}] table[x index=0, y index=1, header=false] {fig/results/var/data/var_hsw_cusp.data};
  \addplot[green, mark=square*, mark options={fill=white}] table[x index=0, y index=3, header=false] {fig/results/var/data/var_hsw_cusp.data};  
  \addplot[blue, mark=triangle*, mark options={fill=white}] table[x index=0, y index=2, header=false] {fig/results/var/data/var_hsw_cusp.data};
  \end{axis}
\end{tikzpicture}}
  \hfill
  \scalebox{.93}{\begin{tikzpicture}[y=.2cm, x=.7cm]
  \centering
  \begin{axis}[
    height=4.5cm,
    width=0.4\textwidth,
    grid=major,
    tick align=inside,
    xlabel={n},
    xmin=0,
    xmax=3.6e+07,
    ymin=3.0e-08,
    ymax=4.0e-08,
  ]
  \addplot[red, mark=*, mark options={fill=white}] table[x index=0, y index=1, header=false] {fig/results/var/data/var_ivy_cusp.data};
  \addplot[green, mark=square*, mark options={fill=white}] table[x index=0, y index=3, header=false] {fig/results/var/data/var_ivy_cusp.data};  
  \addplot[blue, mark=triangle*, mark options={fill=white}] table[x index=0, y index=2, header=false] {fig/results/var/data/var_ivy_cusp.data};
  \end{axis}
\end{tikzpicture}}
  \hfill
  \scalebox{.93}{\begin{tikzpicture}[y=.2cm, x=.7cm]
  \centering
  \begin{axis}[
    height=4.5cm,
    width=0.4\textwidth,
    grid=major,
    tick align=inside,
    xlabel={n},
    xmin=0,
    xmax=3.6e+07,
    ymin=1.1e-08,
    ymax=1.4e-08,
  ]
  \addplot[red, mark=*, mark options={fill=white}] table[x index=0, y index=1, header=false] {fig/results/var/data/var_sky_cusp.data};
  \addplot[green, mark=square*, mark options={fill=white}] table[x index=0, y index=3, header=false] {fig/results/var/data/var_sky_cusp.data};
  \addplot[blue, mark=triangle*, mark options={fill=white}] table[x index=0, y index=2, header=false] {fig/results/var/data/var_sky_cusp.data};
  \end{axis}
\end{tikzpicture}}
  \hfill
  \vspace{.5ex}
  \scalebox{0.93}{\begin{tikzpicture}
  \centering
  \begin{customlegend}[
    legend columns=3,
    legend style={
      /tikz/every even column/.append style={column sep=0.25cm}},
    legend entries={BestVariant,
                    OffsitePreselect5,
                    OffsitePreselect10}]
    \addlegendimage{color=red, thick, mark=*, mark options={fill=white}}
    \addlegendimage{color=blue, thick, mark=triangle*, mark options={fill=white}}
    \addlegendimage{color=green, thick, mark=square*, mark options={fill=white}}
  \end{customlegend}
  \hfill
  \draw (-5.75, -0.25) node[thick, align=center] (hsw) {(a) Haswell (8 cores)};
  \draw (-1.25, -0.25) node[thick, align=center] (ivy) {(b) IvyBridge (10 cores)};
  \draw (2.75, -0.25) node[thick, align=center] (sky) {(c) Skylake (20 cores)};
  \hfill  
\end{tikzpicture}}  
  \vspace{-2.25ex}
  \caption{Comparison of AT strategies applied to \emph{Cusp} with varying $n$ 
  and \emph{Radau II\,A(7)}.}
  \label{fig:at_cusp_var}
  \vspace{-1ex}
\end{figure*}
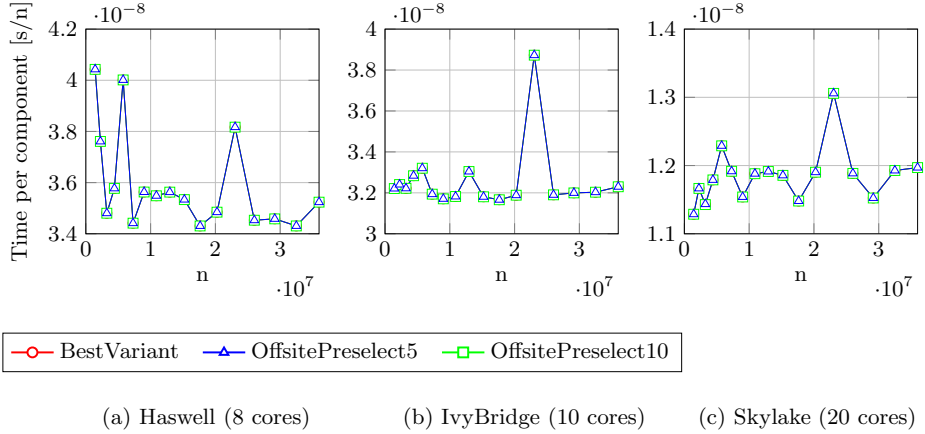

\subsubsection*{Tuning Cusp (Fig. \ref{fig:at_cusp_var}).}
\label{sec:experiments:unknown_n:cusp}


On \emph{Haswell} (Fig. \ref{fig:at_cusp_var}a), \emph{OffsitePreselect5} and 
\emph{OffsitePreselect10} select the same subset of three variants independent 
of $n$. Both strategies always correctly identify the measured best variant.
The same observations can be made on \emph{IvyBridge} (Fig. 
\ref{fig:at_cusp_var}b) and on \emph{Skylake} (Fig. \ref{fig:at_cusp_var}c) 
where also the same subset of three variants is selected and the measured best 
variant is always found.


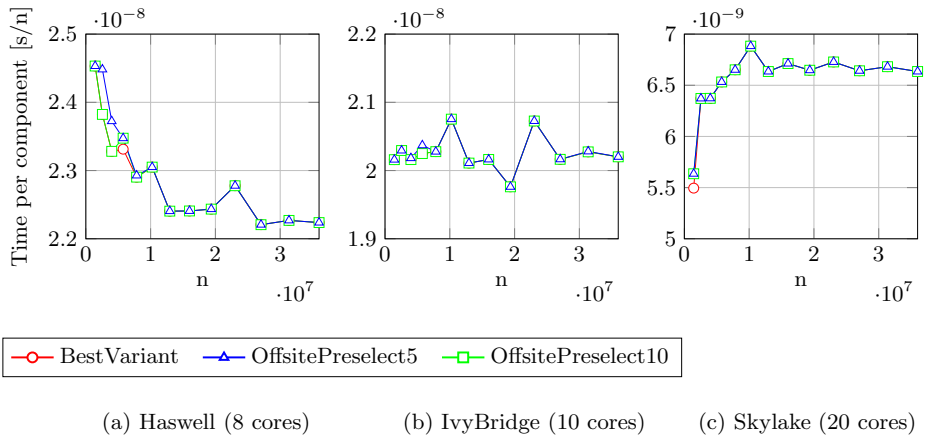
\begin{figure*}[tbp]
  \centering 
  \scalebox{0.93}{


\begin{tikzpicture}[y=.2cm, x=.7cm]
  \centering
  \begin{axis}[
    height=4.5cm,
    width=0.4\textwidth,
    grid=major,
    tick align=inside,
    xlabel={n},
    ylabel={Time per component [s/n]},
    xmin=0,
    xmax=3.6e+07,
    ymin=2.2e-08,
    ymax=2.5e-08,
  ]
  \addplot[red, mark=*, mark options={fill=white}] table[x index=0, y index=1, header=false] {fig/results/var/data/var_hsw_ic.data};
  \addplot[green, mark=square*, mark options={fill=white}] table[x index=0, y index=3, header=false] {fig/results/var/data/var_hsw_ic.data};  
  \addplot[blue, mark=triangle*, mark options={fill=white}] table[x index=0, y index=2, header=false] {fig/results/var/data/var_hsw_ic.data};
  \end{axis}
\end{tikzpicture}}
  \hfill
  \scalebox{0.93}{\begin{tikzpicture}[y=.2cm, x=.7cm]
  \centering
  \begin{axis}[
    height=4.5cm,
    width=0.4\textwidth,
    grid=major,
    tick align=inside,
    xlabel={n},
    xmin=0,
    xmax=3.6e+07,
    ymin=1.9e-08,
    ymax=2.2e-08,
  ]
  \addplot[red, mark=*, mark options={fill=white}] table[x index=0, y index=1, header=false] {fig/results/var/data/var_ivy_ic.data};
  \addplot[green, mark=square*, mark options={fill=white}] table[x index=0, y index=3, header=false] {fig/results/var/data/var_ivy_ic.data};
  \addplot[blue, mark=triangle*, mark options={fill=white}] table[x index=0, y index=2, header=false] {fig/results/var/data/var_ivy_ic.data};
  \end{axis}
\end{tikzpicture}}
  \hfill
  \scalebox{0.93}{\begin{tikzpicture}[y=.2cm, x=.7cm]
  \centering
  \begin{axis}[
    height=4.5cm,
    width=0.4\textwidth,
    grid=major,
    tick align=inside,
    xlabel={n},
    xmin=0,
    xmax=3.6e+07,
    ymin=5.0e-09,
    ymax=7.0e-09,
  ]
  \addplot[red, mark=*, mark options={fill=white}] table[x index=0, y index=1, header=false] {fig/results/var/data/var_sky_ic.data};
  \addplot[green, mark=square*, mark options={fill=white}] table[x index=0, y index=3, header=false] {fig/results/var/data/var_sky_ic.data};  
  \addplot[blue, mark=triangle*, mark options={fill=white}] table[x index=0, y index=2, header=false] {fig/results/var/data/var_sky_ic.data};
  \end{axis}
\end{tikzpicture}}
  \hfill
  \vspace{.5ex}  
  \scalebox{0.93}{\begin{tikzpicture}
  \centering
  \begin{customlegend}[
    legend columns=3,
    legend style={
      /tikz/every even column/.append style={column sep=0.25cm}},
    legend entries={BestVariant,
                    OffsitePreselect5,
                    OffsitePreselect10}]
    \addlegendimage{color=red, thick, mark=*, mark options={fill=white}}
    \addlegendimage{color=blue, thick, mark=triangle*, mark options={fill=white}}
    \addlegendimage{color=green, thick, mark=square*, mark options={fill=white}}
  \end{customlegend}
  \hfill
  \draw (-5.75, -0.25) node[thick, align=center] (hsw) {(a) Haswell (8 cores)};
  \draw (-1.25, -0.25) node[thick, align=center] (ivy) {(b) IvyBridge (10 cores)};
  \draw (2.75, -0.25) node[thick, align=center] (sky) {(c) Skylake (20 cores)};
  \hfill  
\end{tikzpicture}}
  \hfill
  \vspace{-2.25ex}
  \caption{Comparison of AT strategies applied to \emph{IC} with varying $n$ and
  \emph{Radau II\,A(7)}.}
  \label{fig:at_ic_var}
  \vspace{-1ex}
\end{figure*}

\subsubsection*{Tuning IC (Fig. \ref{fig:at_ic_var}).}
\label{sec:experiments:unknown_n:ic}


On \emph{Haswell} (Fig. \ref{fig:at_ic_var}a), the same subset of one (for 
\emph{OffsitePreselect5}) respectively of two variants (for 
\emph{OffsitePreselect10}) is picked for $n$ up to $8{,}500{,}000$. For bigger 
$n$, both strategies select the same three variants. Except for 
$n=5{,}760{,}000$, \emph{OffsitePreselect10} always correctly finds the measured
best variant. The single variant selected by \emph{OffsitePreselect5} is 
slightly off for $n=1{,}440{,}000$ and $n=2{,}560{,}000$. In both cases, 
however, the absolute time difference is only marginal. \emph{IC} is 
compute-bound (Table \ref{tab:ivps}) and, thus, the IVP evaluation dominates the
computation time. Hence, in particular for small $n$, the order of the remaining
computations has only minor impact on the time and already minor jitter can lead
to a different variant being selected.

\emph{OffsitePreslect5} selects on \emph{IvyBridge} (Fig. \ref{fig:at_ic_var}b) 
the same variant for all $n$ while \emph{OffsitePreselect10} adds two additional
variants for $n \ge 2{,}560{,}000$. While \emph{OffsitePreselect10} always finds
the measured best variant, \emph{OffsitePreselect5} is slightly off for 
$n=4{,}000{,}000$ and $n=5{,}760{,}000$ but the absolute time difference is only
marginal.

On Skylake (Fig. \ref{fig:at_ic_var}c), the same variant is selected for $n$ up 
to $1{,}440{,}000$ by \emph{OffsitePreselect5} as well as by 
\emph{OffsitePreselect10} while for larger $n$ two additional variants are 
considered. Except for $n = 1{,}440{,}000$ both Offsite strategies manage to 
always correctly identify the measured best variant. As on the two previous 
systems, the absolute time difference is again only marginal. 


\subsection{AT Scenario -- Variable Number of Cores}
\label{sec:experiments:core}


Offsite is capable of predicting the performance of an implementation variant 
for different core counts with a single AT run. In this AT scenario, we consider
tuning an IVP for a fixed ODE system size $n$ and multiple core counts.

Figure \ref{fig:at_cores} shows the effectiveness of different AT strategies 
compared to strategy \emph{RunAll} when tuning IVP \emph{IC} on three 
target platforms for $n = 9{,}000{,}000$ and \emph{Radau II\,A(7)}. On the 
$x$-axis, we plot the number of cores. The $y$-axis plots for different AT 
strategies the percent \emph{performance gain} $\Pi$ achieved by applying that 
particular strategy instead of \emph{RunAll} which tests all $56$ variants 
($t_\textrm{RA})$. The performance gain is defined as $\frac{t_\textrm{RA} - 
t_\textrm{AT}}{t_\textrm{RA}}*100$ where $t_\textrm{AT}$ includes the time to 
run the variants $\Lambda$ tested by that strategy and the time to run the 
measured best variant from $\Lambda$ an additional $56 - \vert \Lambda \vert$ 
times. Ideally, the bar of an AT strategy would be close to the horizontal line 
of \emph{BestVariant}.



\subsubsection{Haswell (Fig. \ref{fig:at_cores}a).}
\label{sec:experiments:core:haswell}


Depending on the number of cores, \emph{OffsitePreselect5} picks different 
subsets $\Lambda$. For core counts smaller than eight, the same variant is 
selected, while for eight cores two additional variants are selected. Using 
\emph{OffsitePreselect10}, these two variants are also included for four cores. 
For all core counts, a significant performance gain close to 
\emph{BestImplVariant} can be observed when using the Offsite strategies. The 
total performance gain grows with increasing number of cores as the performance gap between best and worst variants also increases. While outperforming \emph{RunAll}, \emph{RandomSelect} is still far off from the maximum gain.


\subsubsection{IvyBridge (Fig. \ref{fig:at_cores}b).}
\label{sec:experiments:core:ivy}


\emph{OffsitePreselct5} selects the same variant for all core counts. Using 
\emph{OffsitePreselect10}, only for 20 cores two further variants are 
selected. Again, a significant performance gain close to \emph{BestImplVariant},
can be observed for all core counts when using the Offsite strategies while 
\emph{RandomSelect} is far off from that ideal gain.



\subsubsection*{Skylake (Fig. \ref{fig:at_cores}c).}
\label{sec:experiments:core:skylake}


Both Offsite strategies select the same three variants for 20 cores, while the 
same single variant is selected for smaller core counts. As on the two previous 
target platforms, both Offsite strategies are close too \emph{BestImplVariant} 
while \emph{RandomSelect} is again further off.



\begin{figure}[tbp]
  \centering
  \scalebox{0.53}{

\begin{tikzpicture}
  \centering
  \pgfplotsset{set layers}
  \begin{axis}[
    ybar,
    scale only axis,
    height=5cm, width=0.5\textwidth,
    bar width=0.22cm,
    ymajorgrids, tick align=inside,
    enlarge y limits={value=.1,upper},
    axis x line*=bottom,
    axis y line*=left,
    y axis line style={opacity=0},
    tickwidth=0pt,
    enlarge x limits=true,
    ylabel={Performance gain [\%]},
    symbolic x coords={1 core, 2 cores, 4 cores, 8 cores},
    xtick=data,
    xlabel style = {font=\large},
    ylabel style = {font=\large},
    xticklabel style = {font=\large},
    yticklabel style = {font=\large},
    ymin=0,
    ymax=60,
  ]
    \addplot [draw=none, fill=blue!50, postaction={pattern=horizontal lines}] coordinates {
      (1 core, 3.72E+01)
      (2 cores, 3.70E+01)
      (4 cores, 3.90E+01)
      (8 cores, 4.67E+01) };
   \addplot [draw=none, fill=green!50, postaction={pattern=dots}] coordinates {
      (1 core, 3.72E+01)
      (2 cores, 3.70E+01) 
      (4 cores, 3.95E+01)
      (8 cores, 4.67E+01) };
   \addplot [draw=none, fill=yellow!50, postaction={pattern=vertical lines}] coordinates {
      (1 core, 2.43E+01)
      (2 cores, 2.31E+01) 
      (4 cores, 2.61E+01)
      (8 cores, 2.65E+01) };
    \addplot[red,sharp plot,update limits=false,] coordinates { ([normalized] -0.5, 3.77E+01) ([normalized] 0.5, 3.77E+01)};
    \addplot[red,sharp plot,update limits=false,] coordinates { ([normalized] 0.5, 3.74E+01) ([normalized] 1.5, 3.74E+01)}; 
    \addplot[red,sharp plot,update limits=false,] coordinates { ([normalized] 1.5, 3.96E+01) ([normalized] 2.5, 3.96E+01)}; 
    \addplot[red,sharp plot,update limits=false,] coordinates { ([normalized] 2.5, 4.68E+01) ([normalized] 3.5, 4.68E+01)};
  \end{axis}
\end{tikzpicture}}
  \hfill
  \scalebox{0.53}{

\begin{tikzpicture}
  \centering
  \pgfplotsset{set layers}
  \begin{axis}[
    ybar,
    scale only axis,
    height=5cm, width=0.5\textwidth,
    bar width=0.22cm,
    ymajorgrids, tick align=inside,
    enlarge y limits={value=.1,upper},
    axis x line*=bottom,
    axis y line*=left,
    y axis line style={opacity=0},
    tickwidth=0pt,
    enlarge x limits=true,
    symbolic x coords={1 core, 2 cores, 5 cores, 10 cores},
    xtick=data,
    xlabel style = {font=\large},
    ylabel style = {font=\large},
    xticklabel style = {font=\large},
    yticklabel style = {font=\large},
    ymin=0,
    ymax=60
  ]
    \addplot [draw=none, fill=blue!50, postaction={pattern=horizontal lines}] coordinates {
      (1 core, 3.54E+01
)
      (2 cores, 3.65E+01
) 
      (5 cores, 4.38E+01
)
      (10 cores, 5.91E+01
) };
   \addplot [draw=none, fill=green!50, postaction={pattern=dots}] coordinates {
      (1 core, 3.54E+01
)
      (2 cores, 3.65E+01
) 
      (5 cores, 4.38E+01
)
      (10 cores, 5.95E+01
) };
   \addplot [draw=none, fill=yellow!50, postaction={pattern=vertical lines}] coordinates {
      (1 core, 2.10E+01
)
      (2 cores, 2.23E+01
) 
      (5 cores, 2.51E+01
)
      (10 cores, 4.07E+01
) };
    \addplot[red,sharp plot,update limits=false,] coordinates { ([normalized] -0.5, 3.54E+01
) ([normalized] 0.5, 3.54E+01
)};
    \addplot[red,sharp plot,update limits=false,] coordinates { ([normalized] 0.5, 3.65E+01
) ([normalized] 1.5, 3.65E+01
)}; 
    \addplot[red,sharp plot,update limits=false,] coordinates { ([normalized] 1.5, 4.39E+01
) ([normalized] 2.5, 4.39E+01
)}; 
    \addplot[red,sharp plot,update limits=false,] coordinates { ([normalized] 2.5, 5.95E+01) ([normalized] 3.5, 5.95E+01
)};  
  \end{axis}
\end{tikzpicture}}
  \hfill  
  \scalebox{0.53}{

\begin{tikzpicture}
  \centering
  \pgfplotsset{set layers}
  \begin{axis}[
    ybar,
    scale only axis,
    height=5cm, width=0.5\textwidth,
    bar width=0.22cm,
    ymajorgrids, tick align=inside,
    enlarge y limits={value=.1,upper},
    axis x line*=bottom,
    axis y line*=left,
    y axis line style={opacity=0},
    tickwidth=0pt,
    enlarge x limits=true,
    symbolic x coords={1 core, 5 cores, 10 cores, 20 cores},
    xtick=data,
    xlabel style = {font=\large},
    ylabel style = {font=\large},
    xticklabel style = {font=\large},
    yticklabel style = {font=\large},
    ymin=0,
    ymax=60
  ]
    \addplot [draw=none, fill=blue!50, postaction={pattern=horizontal lines}] coordinates {
      (1 core, 4.67E+01)
      (5 cores, 4.92E+01) 
      (10 cores, 6.10E+01)
      (20 cores, 6.39E+01) };
   \addplot [draw=none, fill=green!50, postaction={pattern=dots}] coordinates {
      (1 core, 4.66E+01)
      (5 cores, 4.92E+01) 
      (10 cores, 6.10E+01)
      (20 cores, 6.39E+01) };
   \addplot [draw=none, fill=yellow!50, postaction={pattern=vertical lines}] coordinates {
      (1 core, 2.87E+01)
      (5 cores, 3.30E+01) 
      (10 cores, 4.08E+01)
      (20 cores, 3.86E+01) };
    \addplot[red,sharp plot,update limits=false,] coordinates { ([normalized] -0.5, 4.74E+01) ([normalized] 0.5, 4.74E+01)};
    \addplot[red,sharp plot,update limits=false,] coordinates { ([normalized] 0.5, 5.03E+01) ([normalized] 1.5, 5.03E+01)}; 
    \addplot[red,sharp plot,update limits=false,] coordinates { ([normalized] 1.5, 6.11E+01) ([normalized] 2.5, 6.11E+01)}; 
    \addplot[red,sharp plot,update limits=false,] coordinates { ([normalized] 2.5, 6.39E+01) ([normalized] 3.5, 6.39E+01)};
  \end{axis}
\end{tikzpicture}}
  \hfill
  \scalebox{0.75}{\begin{tikzpicture}
  \centering
  \begin{customlegend}[
    legend columns=4,
    legend style={
      anchor=north,
      /tikz/every even column/.append style={column sep=0.25cm}},
    legend entries={BestVariant,
                    OffsitePreselect5,
                    OffsitePreselect10,
                    RandomSelect}]
    \addlegendimage{color=red!50, font=\normalsize}                    
    \addlegendimage{color=blue!50, area legend, pattern=horizontal lines, font=\normalsize}
    \addlegendimage{color=green!50, area legend, pattern=dots, font=\normalsize}
    \addlegendimage{color=yellow!50, area legend, pattern=vertical lines, font=\normalsize}
  \end{customlegend}
  \hfill
  \draw (-3.0, -0.25) node[thick, align=center, font=\normalsize] (hsw) {(a) Haswell};
  \draw ( 2.5, -0.25) node[thick, align=center, font=\normalsize] (ivy) {(b) IvyBridge};
  \draw ( 7.5, -0.25) node[thick, align=center, font=\normalsize] (sky) {(c) Skylake};
  \hfill
\end{tikzpicture}

}
  \vspace{-2.25ex}  
  \caption{Percent performance gain achieved by different AT strategies when 
  tuning IVP \emph{IC} for different core counts, \emph{Radau II\,A(7)} and $n =
  9{,}000{,}000$. }
  \label{fig:at_cores}
  \vspace{-1.0ex}    
\end{figure}


\subsection{AT Scenario -- Variable ODE Method}
\label{sec:experiments:ode}


In the last AT scenario, we consider tuning an IVP for a fixed ODE system 
size $n$ for four different ODE methods. Depending on the characteristics of the
ODE method, different optimizations might be applicable---for specific number of
stages $s$, e.g., loops over $s$ can be replaced by a vector operation---which 
potentially results in varying efficiency of the same implementation variant for
different ODE methods.

Figure \ref{fig:at_methods} shows the effectiveness of different AT strategies 
when tuning IVP \emph{IC} on three target platforms for $n = 9{,}000{,}000$ and 
four different ODE methods: \emph{(i)} \emph{Radau I\,A(5)} ($s=3$, $m=4$), 
\emph{(ii)} \emph{Radau II\,A(7)} ($s=4$, $m=6$), \emph{(iii)} \emph{Lobatto 
III\,C(6)} ($s=4$, $m=5$), and \emph{(iv)}~\emph{Lobatto III\,C(8)} ($s=5$, 
$m=7$).
On the $x$-axis the ODE method used is shown. The $y$-axis plots for each AT 
strategy the percent performance gain $\Pi$ achieved by applying that  
particular strategy instead of \emph{RunAll} which tests all $56$ variants. The 
bar of an AT strategy is ideally close to the horizontal line of 
\emph{BestVariant}.


\subsubsection{Tuning Haswell (Fig. \ref{fig:at_methods}a).}
\label{sec:experiments:method:haswell}


\emph{OffsitePreselect5} selects the same subset of two variants for 
\emph{Lobatto III\,C(6)} and \emph{Radau I\,A(5)}. For \emph{Lobatto III\,C(8)} 
and \emph{Radau  II\,A(7)}, an additional variant is selected. Using 
\emph{OffsitePreselect10}, these three variants are selected for all ODE 
methods.
For all ODE methods, a significant performance gain close to 
\emph{BestImplVariant} can be observed when using one of the two Offsite 
strategies. Further, both Offsite strategies decisively outperform 
\emph{RandomSelect}.


\subsubsection{IvyBridge (Fig. \ref{fig:at_methods}b).}
\label{sec:experiments:method:ivy}


For all ODE methods, the same single variant is chosen when using 
\emph{OffsitePreselect5}, while \emph{OffsitePreselect10} selects two variants 
for \emph{Lobatto III\,C(6)} and the same three variants for \emph{Lobatto 
III\,C(8)} and \emph{Radau  II\,A(7)}. As on \emph{Haswell}, the performance 
gain of both Offsite strategies for all ODE methods is close to the maximum 
gain, while the achieved gain of \emph{RandomSelect} is far off from 
\emph{BestImplVariant}.

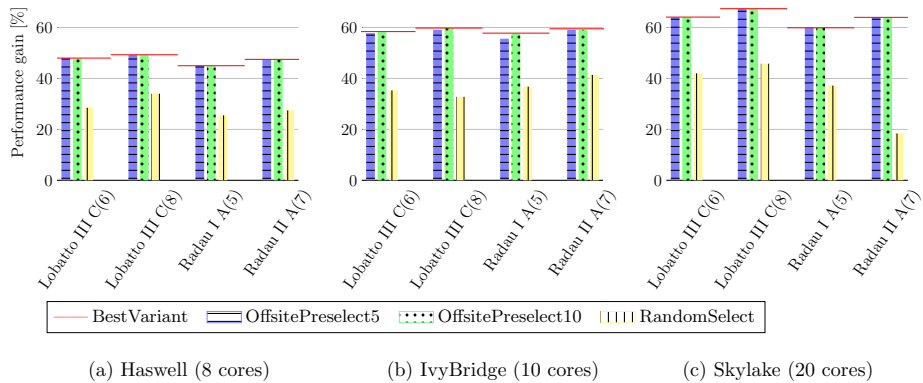
\begin{figure}[tbp]
  \centering
  \scalebox{0.52}{

\begin{tikzpicture}
  \centering
  \pgfplotsset{set layers}
  \begin{axis}[
    ybar,
    scale only axis,
    height=5cm, width=0.5\textwidth,
    bar width=0.22cm,
    ymajorgrids, tick align=inside,
    enlarge y limits={value=.1,upper},
    axis x line*=bottom,
    axis y line*=left,
    y axis line style={opacity=0},
    tickwidth=0pt,
    enlarge x limits=true,
    ylabel={Performance gain [\%]},
    symbolic x coords={Lobatto III C(6), Lobatto III C(8), Radau I A(5), Radau II A(7)},
    x tick label style = {rotate=55},
    xtick=data,
    xlabel style = {font=\large},
    ylabel style = {font=\large},
    xticklabel style = {font=\large},
    yticklabel style = {font=\large},
    ymin=0,
    ymax=70
  ]
    \addplot [draw=none, fill=blue!50, postaction={pattern=horizontal lines}] coordinates {
      (Lobatto III C(6), 4.80E+01)
      (Lobatto III C(8), 4.92E+01) 
      (Radau I A(5), 4.49E+01)
      (Radau II A(7), 4.74E+01) };
  \addplot [draw=none, fill=green!50, postaction={pattern=dots}] coordinates {
      (Lobatto III C(6), 4.79E+01)
      (Lobatto III C(8), 4.92E+01) 
      (Radau I A(5), 4.48E+01)
      (Radau II A(7), 4.74E+01) };
  \addplot [draw=none, fill=yellow!50, postaction={pattern=vertical lines}] coordinates {
      (Lobatto III C(6), 2.85E+01)
      (Lobatto III C(8), 3.40E+01) 
      (Radau I A(5), 2.55E+01)
      (Radau II A(7), 2.75E+01) };
    \addplot[red,sharp plot,update limits=false,] coordinates { ([normalized] -0.5, 4.80E+01) ([normalized] 0.5, 4.80E+01)};
    \addplot[red,sharp plot,update limits=false,] coordinates { ([normalized] 0.5, 4.93E+01) ([normalized] 1.5, 4.93E+01)}; 
    \addplot[red,sharp plot,update limits=false,] coordinates { ([normalized] 1.5, 4.50E+01) ([normalized] 2.5, 4.50E+01)}; 
    \addplot[red,sharp plot,update limits=false,] coordinates { ([normalized] 2.5, 4.75E+01) ([normalized] 3.5, 4.75E+01)};
  \end{axis}
\end{tikzpicture}}
  \hfill
  \scalebox{0.52}{\begin{tikzpicture}
  \centering
  \pgfplotsset{set layers}
  \begin{axis}[
    ybar,
    scale only axis,
    height=5cm, width=0.5\textwidth,
    bar width=0.22cm,
    ymajorgrids, tick align=inside,
    enlarge y limits={value=.1,upper},
    axis x line*=bottom,
    axis y line*=left,
    y axis line style={opacity=0},
    tickwidth=0pt,
    enlarge x limits=true,
    symbolic x coords={Lobatto III C(6), Lobatto III C(8), Radau I A(5), Radau II A(7)},
    x tick label style = {rotate=55},
    xtick=data,
    xlabel style = {font=\large},
    ylabel style = {font=\large},
    xticklabel style = {font=\large},
    yticklabel style = {font=\large},
    ymin=0,
    ymax=70
  ]
    \addplot [draw=none, fill=blue!50, postaction={pattern=horizontal lines}] coordinates {
      (Lobatto III C(6), 5.78E+01)
      (Lobatto III C(8), 5.91E+01) 
      (Radau I A(5), 5.56E+01)
      (Radau II A(7), 5.91E+01) };
  \addplot [draw=none, fill=green!50, postaction={pattern=dots}] coordinates {
      (Lobatto III C(6), 5.84E+01)
      (Lobatto III C(8), 5.97E+01) 
      (Radau I A(5), 5.77E+01)
      (Radau II A(7), 5.91E+01) };
  \addplot [draw=none, fill=yellow!50, postaction={pattern=vertical lines}] coordinates {
      (Lobatto III C(6), 3.53E+01)
      (Lobatto III C(8), 3.28E+01) 
      (Radau I A(5), 3.68E+01)
      (Radau II A(7), 4.14E+01) };
    \addplot[red,sharp plot,update limits=false,] coordinates { ([normalized] -0.5, 5.84E+01) ([normalized] 0.5, 5.84E+01)};
    \addplot[red,sharp plot,update limits=false,] coordinates { ([normalized] 0.5, 5.98E+01) ([normalized] 1.5,5.98E+01)}; 
    \addplot[red,sharp plot,update limits=false,] coordinates { ([normalized] 1.5, 5.78E+01) ([normalized] 2.5, 5.78E+01)}; 
    \addplot[red,sharp plot,update limits=false,] coordinates { ([normalized] 2.5, 5.95E+01) ([normalized] 3.5, 5.95E+01)};
  \end{axis}
\end{tikzpicture}}
  \hfill  
  \scalebox{0.52}{\begin{tikzpicture}
  \centering
  \pgfplotsset{set layers}
  \begin{axis}[
    ybar,
    scale only axis,
    height=5cm, width=0.5\textwidth,
    bar width=0.22cm,
    ymajorgrids, tick align=inside,
    enlarge y limits={value=.1,upper},
    axis x line*=bottom,
    axis y line*=left,
    y axis line style={opacity=0},
    tickwidth=0pt,
    enlarge x limits=true,
    symbolic x coords={Lobatto III C(6), Lobatto III C(8), Radau I A(5), Radau II A(7)},
    x tick label style = {rotate=55},
    xtick=data,
    xlabel style = {font=\large},
    ylabel style = {font=\large},
    xticklabel style = {font=\large},
    yticklabel style = {font=\large},
    ymin=0,
    ymax=70
  ]
    \addplot [draw=none, fill=blue!50, postaction={pattern=horizontal lines}] coordinates {
      (Lobatto III C(6), 6.40E+01)
      (Lobatto III C(8), 6.73E+01) 
      (Radau I A(5), 5.99E+01)
      (Radau II A(7), 6.40E+01) };
   \addplot [draw=none, fill=green!50, postaction={pattern=dots}] coordinates {
      (Lobatto III C(6), 6.40E+01)
      (Lobatto III C(8), 6.73E+01) 
      (Radau I A(5), 5.99E+01)
      (Radau II A(7), 6.40E+01) };
   \addplot [draw=none, fill=yellow!50, postaction={pattern=vertical lines}] coordinates {
      (Lobatto III C(6), 4.20E+01)
      (Lobatto III C(8), 4.58E+01) 
      (Radau I A(5), 3.72E+01)
      (Radau II A(7), 1.84E+01) };
    \addplot[red,sharp plot,update limits=false,] coordinates { ([normalized] -0.5, 6.41E+01) ([normalized] 0.5, 6.41E+01)};
    \addplot[red,sharp plot,update limits=false,] coordinates { ([normalized] 0.5, 6.74E+01
) ([normalized] 1.5, 6.74E+01
)}; 
    \addplot[red,sharp plot,update limits=false,] coordinates { ([normalized] 1.5, 5.99E+01) ([normalized] 2.5, 5.99E+01)}; 
    \addplot[red,sharp plot,update limits=false,] coordinates { ([normalized] 2.5, 6.40E+01) ([normalized] 3.5, 6.40E+01)};
  \end{axis}
\end{tikzpicture}}
  \hfill  
  \scalebox{0.75}{\begin{tikzpicture}
  \centering
  \begin{customlegend}[
    legend columns=4,
    legend style={
      anchor=north,
      /tikz/every even column/.append style={column sep=0.25cm}},
    legend entries={BestVariant,
                    OffsitePreselect5,
                    OffsitePreselect10,
                    RandomSelect}]
    \addlegendimage{color=red!50, font=\normalsize}                    
    \addlegendimage{color=blue!50, area legend, pattern=horizontal lines, font=\normalsize}
    \addlegendimage{color=green!50, area legend, pattern=dots, font=\normalsize}
    \addlegendimage{color=yellow!50, area legend, pattern=vertical lines, font=\normalsize}
  \end{customlegend}
  \hfill
  \draw (-3.0, -0.25) node[thick, align=center, font=\normalsize] (ivy) {(a) Haswell (8 cores)};
  \draw ( 2.5, -0.25) node[thick, align=center, font=\normalsize] (ivy) {(b) IvyBridge (10 cores)};
  \draw ( 7.5, -0.25) node[thick, align=center, font=\normalsize] (ivy) {(c) Skylake (20 cores)};
  \hfill  
\end{tikzpicture}

}
  \vspace{-2.25ex}
  \caption{Percent performance gain achieved by different AT strategies when 
  tuning IVP \emph{IC} for different ODE methods and $n = 9{,}000{,}000$. }  
  \label{fig:at_methods}
  \vspace{-1.0ex}  
\end{figure}

\subsubsection{Skylake (Fig. \ref{fig:at_methods}c).}
\label{sec:experiments:method:skylake}


Both Offsite AT strategies select the same subset of three variants for all ODE 
methods but for \emph{Radau I\,A(5)} which only selects two variants when using
\emph{OffsitePreselect5}. Again, the performance gain achieved by both Offsite 
strategies is close to \emph{BestVariant} while \emph{RandomSearch} is further 
off.


\section{Conclusion and Future Work}
\label{sec:conclusion}


In this work, we have introduced the Offsite AT approach which automates the 
process of identifying the most efficient implementation variant(s) from a pool 
of possible variants at installation time. Offsite ranks variants by their 
performance using analytic performance predictions. To facilitate specifying 
tuning scenarios, multilevel YAML description languages allow to describe these 
scenarios in an abstract way and enable Offsite to automatically generate 
optimized codes. Moreover, we have demonstrated that Offsite can reliably tune a
representative class of parallel explicit ODE methods, PIRK methods, by 
investigating different AT scenarios and AT strategies on three different 
shared-memory platforms.

Our future work includes expanding Offsite to cluster systems as well as to
AMD and ARM platforms. Further, we intend to extend Offsite to a combined 
off\/line-online AT approach that incorporates feedback data from previous online
AT (or program runs)
and to study whether these data can be used to predict the performance in scenarios with unknown input data 
(e.g. new IVP).


\section*{Acknowledgments}


This work is supported by the German Ministry of Science and Education (BMBF)
under project number 01IH16012A.
Furthermore, we thank the Erlangen Regional Computing Center (RRZE) for granting
access to their IvyBridge and Skylake systems.

\bibliographystyle{splncs04}
\bibliography{paper}


\end{document}